\documentclass{article}
\usepackage{graphicx}
\usepackage{here}

\newcommand{\AmS}{{\protect\the\textfont2
  A\kern-.1667em\lower.5ex\hbox{M}\kern-.125emS}}
\def\beq{\begin{equation}}
\def\eeq{\end{equation}}
\def\bea{\begin{eqnarray}}
\def\eea{\end{eqnarray}}
\def\bq{\begin{quote}}
\def\eq{\end{quote}}

\def\bear{\begin{array}}
\def\ear{\end{array}}
\def\nnb{\nonumber}
\def\ga{\left(}
\def\dr{\right)}

\def\rar{\rightarrow}

\def\nnb{\nonumber}

\def\nin{\noindent}
\def\ba{\begin{array}}
\def\ea{\end{array}}

\def\gam5{\gamma_5}

\begin{document}
\begin{center}
\section*{{Probing High-Energy Physics with High-precision QED measurements}
\footnote{Talk presented
 at the 1st High-Energy
Physics Madagascar International Conference Series (HEP-MAD'01), 27th sept-5th Oct. 2001,
Antananarivo (to be published by World Scientific, Singapore).}}

\vspace*{1.5cm}
{\bf Stephan Narison} \\
\vspace{0.3cm}
Laboratoire de Physique Math\'ematique\\
Universit\'e de Montpellier II\\
Place Eug\`ene Bataillon\\
34095 - Montpellier Cedex 05, France\\
Email: qcd@lpm.univ-montp2.fr\\
\vspace*{1.5cm}
{\bf Abstract} \\ \end{center}
\vspace*{2mm}
\noindent{I summarize our {\it self-contained} determinations of the
lowest order hadronic contributions
\cite{SN,SN2} to the anomalous magnetic moments $a_{\mu,\tau}$ of the muon and tau
leptons, the running QED coupling $\alpha(M_Z)$ and the muonium hyperfine splitting $\nu$. 
Using an average estimate of the light-by light scattering contribution: $a_\mu(LL)=85(18)\times 10^{-11}$,
we deduce:
$a_\mu^{SM} =116~591~861(78)\times 10^{-11}$,
$a_\tau^{SM} =117~759(7)\times 10^{-8}$, giving: $
a_\mu^{SM}-a_\mu^{exp}=162(170)\times 10^{-11}$. We also obtain:
$\alpha^{-1}(M_Z)=128.926(25)$
and the Fermi energy splitting: $\nu_F^{SM}=4~459~031~783(229)~{\rm
Hz}$. Lower bounds
on some new physics are given, while
$\nu_F^{SM}$ leads e.g. to
$m_\mu/m_e=206.768~276(11)$ in remarkable agreement with the data.}
\section{Introduction}
QED is at present the gauge theory where perturbative calculations are the most precise known today.
Therefore, accurate measurements of QED processes are expected to give strong constraints 
on different electroweak
models and to reveal some eventual deviations from the standard model (SM) predictions \footnote{For 
general discussions on astroparticle physics and supersymmetric models, see e.g. \cite{ELLIS,PERAZZI}.}.
In the following, I will discuss the effects of the hadronic and QCD contributions to three classical
QED processes which are: the anomalous magnetic moment of the muon and tau leptons, the running QED
coupling
$\alpha(M_Z)$ and the muonium hyperfine splitting $\nu$. These hadronic contributions are one of the
main sources of uncertainties into these processes. Using a dispersion relation, it is remarkable to
notice that the different  lowest order hadronic contributions for these three processes can be
expressed in a closed form as a convolution of the
$e^+e^-\rar$ hadrons cross-section $\sigma_H(t)$ with a QED kernel function
$K(t)$ which depends on each observable:

\bea
{\cal O}_{\rm had}=\frac{1}{4\pi^3}\int_{4m^2_\pi}^\infty dt~K_{\cal O}(t)~\sigma_H(t)~,
\eea
where:
\beq
{\cal O}_{\rm had}~\equiv~
a_{l,\rm had}~,~~~\Delta\alpha_{\rm had}\times 10^5~~~{\rm or}~~~
\Delta\nu_{\rm had}~.
\eeq
\begin{itemize}
\item~For the anomalous magnetic moment $a_{l,\rm had}$, $K_{a_l}(t\geq 0)$ is
the well-known kernel function \cite{GOURDIN}:
\bea\label{kernel}
K_{a_l}(t)&=&\int_0^1 dx\frac{x^2(1-x)}{x^2+\ga{t}/{m_l^2}\dr(1-x)}~,
\eea
where $m_l$ is the lepton mass. It behaves for large  $t$ as:
\beq
K_{a_l}(t\gg m_l^2)\simeq \frac{m^2_l}{3t}~.
\eeq
\item~For the QED running coupling $\Delta\alpha_{\rm had}\times 10^5$, the
kernel is (see e.g. \cite{JEGER}):
\beq
K_\alpha(t)=\ga\frac{\pi}{\alpha}\dr\ga\frac{M^2_Z}{ M^2_Z-t}\dr~,
\eeq
where $\alpha^{-1}(0)=137.036$ and $M_Z=91.3$ GeV. It behaves for large  $t$
like a constant.
\item~For the muonium hyperfine splitting $\Delta\nu_{\rm had}$, the kernel
function is (see e.g \cite{HFS}):
\beq
K_\nu=-\rho_\nu\Bigg{[}\ga
{x_\mu}+2\dr\ln\frac{1+v_\mu}{1-v_\mu}-\ga
{x_\mu}+\frac{3}{2}\dr\ln x_\mu\Bigg{]}
\eeq
where:
\beq
\rho_\nu= 2\nu_F\frac{m_e}{m_u}~,~~~~~x_\mu=\frac{t}{4m_\mu^2}~~~~~v_\mu=\sqrt{1-\frac{1}{x_\mu}}~,
\eeq
and we take (for the moment) for a closed comparison with \cite{HFS} \footnote{In the
next section, we shall extract this value from the analysis.}, the value of the Fermi energy splitting:
\beq\label{eqhfs}
\nu_F=445~903~192~0.(511)(34)~{\rm Hz}~.
\eeq
It behaves for large  $t$ as:
\beq
K_{\nu}(t\gg m_\mu^2)\simeq \rho_\nu\ga\frac{m^2_\mu}{t}\dr\ga\frac{9}{2}\ln
\frac{t}{m^2_\mu}+\frac{15}{4}\dr~.
\eeq
The different asymptotic behaviours of these kernel functions will influence on
the relative weights of different regions contributions in the evaluation of the above integrals.
\end{itemize}
\section{Input and Numerical Strategy}
\nin
 The
different data input and QCD parametrizations  of the cross-section
$\sigma_H(t)$ have been discussed in details in \cite{SN} (herereferred as SN1)  and corresponding
discussions will not be repeated here. The sources of these data are quoted in the last column of Table 1
from SN1 and Table 2 from \cite{SN2} (herereferred as SN2) are classified according to the estimate in
different regions.  We shall only sketched briefly the numerical strategy here:
\begin{itemize}
\item~Our result from the $I=1$
isovector channel below 3 GeV$^2$ is the mean value of the one using
$\tau$-decay and
$e^+e^-$ data. In both cases, we have used standard trapezo\"\i dal rules and/or least
square fits of the data in order to avoid theoretical model dependence parametrization of the pion form
factor. Correlations among different data have been taken in the compilations of \cite{DAVIE}
used in this paper. In the region
$(0.6-0.8)$ GeV$^2$ around the
$\omega$-$\rho$ mixing, we use in both cases $e^+e^-$ data in order to take
properly the $SU(2)_F$ mixing. The $SU(2)$ breaking in the remaining regions
are taken into account by making the average of the two results from $\tau$-decay
and $e^+e^-$ and by
adding into the errors the distance between this mean central value with the one from
each data.
\item~For the $I=0$ isoscalar channel below 3 GeV$^2$, we use the contributions
of the resonances $\omega(782)$ and $\phi(1020)$ using narrow width
approximation (NWA). We add to these contributions, the
sum of the exclusive channels from 0.66 to 1.93 GeV$^2$. Above 1.93 GeV$^2$, we include
the contributions of the $\omega(1.42),~\omega(1.65)$ and
$\phi(1.68)$ using a Breit-Wigner form of the resonances. 
\item~For the heavy quarkonia, we include the contributions of known $J/\psi$
(1S to 4.415) and $\Upsilon$ (1S to 11.02) families and use a NWA. We have added the effect of the
$\bar tt$ bound state using the leptonic width of $(12.5\pm 1.5)$ keV given in \cite{YND}.
\item~Away from thresholds, we use perturbative QCD plus negligible quark and gluon
condensate contributions, which is expected to give a good parametrization
of the cross-section. These different expressions are given in SN1. However, as the relative r\^ole of
the QCD continuum is important in the estimate of $\Delta\alpha_{\rm had}$, we have added, to the usual
Schwinger interpolating factor at order $\alpha_s$ for describing the heavy quark spectral function, the
known $\alpha_s^2m_Q^2/t$ corrections given in SN1. However, in the region we are working, these
corrections are tiny.
\item~On the $Z$-mass, the integral for $\Delta\alpha_{\rm had}$ has a pole, such that this contribution
has been separated in this case from the QCD continuum. Its value comes from the Cauchy principal
value of the integral.
\end{itemize}
\section{Lowest order hadronic contributions}

\subsection{Muon and tau anomalies}
We show in Table 1 the details of the different hadronic contributions
from each channels and from different energy regions for the muon and tau anomalies. 
Taking the average of the results in Table 1 and adding further systematics due to an eventual deviation
from the CVC assumption and from the choice of the QCD continuum threshold for the light flavours, one
deduce the final estimate from an average of the $e^+e^-$ and $\tau$-decay data \cite{SN}:
\bea\label{final}
a_\mu^{had}(l.o)&=&7020.6(75.6)\times 10^{-11}~,~~
a_\tau^{had}(l.o)=353.6(4.0)\times 10^{-8}~,
\eea
\begin{itemize}
\item The
main error (80\% when added quadratically) in our previous determinations comes from the $\rho$-meson
region below 0.8 GeV$^2$. Hopefully, improved
measurements of this region are feasible in the near future.
\item The second source of errors 
comes from the region around
$M_\tau$ for the inclusive $\tau$-decay and between 1 GeV to $M_{\tau}$  for the $e^+e^-$ data. These
errors are about half of the one from the region below 0.8 GeV$^2$ in most different determinations.
They can be reduced by improved measurements of inclusive $\tau$-decay near $M_\tau$ $(I=1)$ and by
improving the measurements of the odd multi-pions and $\bar KK, ~\bar KK\pi,...$ channels in
the $I=0$ channels from $e^+e^-$ data.
\item The
contributions of the whole region above 3 GeV$^2$ induce much smaller errors (7\% of the total).
There is a quite good consensus between different determinations in this energy region.
\item These predictions agree within the errors with previous predictions quoted in SN1 \cite{SN} and
recent estimates given in
\cite{JEGER}--\cite{PALOMA}.
\end{itemize}
\subsection{Running QED coupling}
Using the same data as for the anomalous magnetic moment, one can deduce from Table 2:
\bea\label{runalfa}
\Delta\alpha_{\rm had}&=&2763.4(16.5)\times 10^{-5}~.
\eea
Also a detailed comparison of each region of energy with the most recent
work of \cite{YND} shows the same features (agreement and slight difference) like in the case of
$a_\mu$ in SN1  due to the slight difference in the
parametrization of the data and spectral function.
However, the final results are comparable. Finally, one
can remark that due to the high-energy constant
behaviour of the QED kernel function in this case, the
low-energy region is no longer dominating. 
 For $a_\mu$, the contribution of the
$\rho$-meson below 1 GeV is 68\% of the total contribution, while the sum of the QCD continuum is only
7.4\% (see e.g. SN1). Here the situation is almost reversed: the contribution of the
$\rho$-meson below 1 GeV is only 2\%, while the sum of the QCD-continuum is 73.6\%. 
\small{\begin{table*}[H]
\setlength{\tabcolsep}{.pc}
\catcode`?=\active \def?{\kern\digitwidth}
\begin{center}
\caption{Determinations of $a_l^{had}(l.o)$ using combined $e^+e^-$
and inclusive $\tau$ decay data (2nd and 4th columns) and averaged $e^+e^-$ data (3rd column).}
\begin{tabular*}{\textwidth}{@{}l@{\extracolsep{\fill}}llll}
\hline
&\\
\multicolumn{1}{l}{\bf Region in GeV$\bf ^2$}
 & \multicolumn{2}{c}{$\bf a_\mu^{had}(l.o)\times 10^{11}$}
  & \multicolumn{1}{l}{$\bf a_\tau^{had}(l.o)\times 10^{8}$}
& \multicolumn{1}{c}{\bf Data input} \\
&\\
\hline\\
                 & \multicolumn{1}{l}{\bf ${\bf\tau}$+$\bf e^+e^-$}
& \multicolumn{1}{l}{$\bf e^+e^-$} 
                 & \multicolumn{1}{l}{\bf ${\bf\tau}$+$\bf e^+e^-$}\\ 
{\bf Light Isovector}&&&\\
$4m_\pi^2\rar 0.8$&$4794.6\pm 60.7$&$4730.2\pm 99.9$&$165.8\pm 1.5$&\cite{DAVIE,ALEPH,OPAL}\\
$0.8\rar 2.1$&$494.9\pm 15.8$&$565.0\pm 54.0$&$28.7\pm 1.1$&\cite{ALEPH,OPAL}\\
$2.1\rar 3.$&$202.0\pm 29.7$&$175.9\pm 16.0$&$17.0\pm 2.6$&\cite{ALEPH,OPAL}\\
\it Total Light I=1&$\it 5491.5\pm 69.4$&$\it 5471.1\pm 114.7$&$\it 211.5\pm 3.2$\\
\bf Light Isoscalar&&\\
{\it Below 1.93 }&\\
$\omega$&$387.5\pm 13$&$387.5\pm 13$&$15.3\pm 0.5$&NWA \cite{PDG}\\
$\phi$&$393.3\pm 9.9$&$393.3\pm 9.9$&$21.0\pm 0.5$&NWA \cite{PDG}\\
$0.66\rar 1.93$&$79.3\pm 18.7$&$79.3\pm 18.7$&$4.3\pm 1.1$&$\sum{\rm exclusive}$ \cite{DOL}\\
{\it From 1.93 to 3~} &\\
$\omega(1.42),~\omega(1.65)$&$31.3\pm 6.8$&$31.3\pm 6.8$&$2.6\pm 0.7$&BW \cite{DM2,PDG}\\
$\phi(1.68)$&$42.4\pm 18.2$&$42.4\pm 18.2$&$3.8\pm 1.3$&BW \cite{DM2,DM1,PDG}\\
{\it Total Light I=0 }&$\it 933.8\pm 31.5$ &$\it 933.8\pm 31.5$&$\it 47.0\pm 2.0$&\\
\bf Heavy Isoscalar&&\\
$J/\psi(1S\rar 4.415)$&$87.0\pm 4.7$&$87.0\pm 4.7$&$13.08\pm 0.69$&NWA \cite{PDG}\\
$\Upsilon(1S\rar 11.020)$&$0.95\pm 0.04$&$0.95\pm 0.04$&$0.23\pm 0.01$&NWA \cite{PDG}\\
\it Total Heavy I=0&$\it 88.0\pm 4.7$&$\it 88.0\pm 4.7$&$\it 13.3\pm 0.7$\\
\bf QCD continuum&&\\
$3.\rar (4.57)^2$&$407.0\pm 2.3$&$407.0\pm 2.3$&$49.4\pm 0.3$&$(u,d,s)$ \\
$(4.57)^2\rar (11.27)^2$&$95.3\pm 0.5$&$95.3\pm 0.5$&$27.3\pm 0.1$&$(u,d,s,c)$ \\
$(11.27)^2\rar 4M^2_t$&$20.5\pm 0.1$&$20.5\pm 0.1$&$5.87\pm 0.01$&$(u,d,s,c,b)$ \\
$4M^2_t\rar \infty$&$\approx 0.$&$\approx 0.$&$\approx 0.$&$(u,d,s,c,b,t)$\\
\it Total QCD Cont.&$\it 522.8\pm 2.4$&$\it 522.8\pm 2.4$&$\it 82.6\pm 0.3$&\\
&\\
\hline
&\\
&7036.1(76.4)&7015.7(119.1)&354.4(3.8)&\\
&\\
\hline\\
\end{tabular*}
\end{center}
\end{table*}}
\nin
For this reason,
improvement due to the new Novosibirsk $e^+e^-$ data \cite{NOVO} in the low-energy region is not 
significant, as we have explicitly checked. At present, new BES data
\cite{BES} in the
$J/\psi$ region are also available, which can be alternatively used.  Below the $J/\psi$ resonances, the BES
data are in excellent agreement with the QCD parametrization to order $\alpha_s^3$ used here for 3 flavours,
justifying the accuracy of your input. Above the $J/\psi$ resonances, the parametrization used here (sum of
narrow resonances +QCD continuum away from thresholds) can also be compared with these data. On can
notice that, in the resonance regions, the BES data are more accurate than previous ones, which may
indicate that our quoted errors in Table 1 for the $J/\psi$ family contributions are overestimated. In
addition, the threshold of the QCD continuum which we have taken above the $J/\psi$ resonances, matches
quite well with the one indicated by the BES data. Our estimate of $\Delta\alpha_{\rm had}$ is compared with
recent predictions respectively in Fig. 1 from SN2 \cite{SN2} (references to the authors are in
\cite{JEGER,YND,DAVIE,ALFA}) where one can notice a very good agreement.

\small{\begin{table*}[H]
\setlength{\tabcolsep}{0.pc}
\newlength{\digitwidth} \settowidth{\digitwidth}{\rm 0}
\catcode`?=\active \def?{\kern\digitwidth}
\caption{Lowest order determinations of $\Delta\alpha_{\rm had}\times 10^{5}$ and
$\Delta\nu_{\rm had}~\rm{[Hz]}$ using combined $e^+e^-$
and inclusive $\tau$ decay data (2nd and 4th columns) and averaged $e^+e^-$ data (3rd and 5th columns).}
\begin{tabular*}{\textwidth}{@{}l@{\extracolsep{\fill}}lllll}
\hline
&\\
\multicolumn{1}{l}{\bf Region in GeV$\bf ^2$}
 & \multicolumn{2}{c}{$\bf \Delta\alpha_{\rm had}\times 10^{5}$}
  & \multicolumn{2}{c}{$\bf \Delta\nu_{\rm had}~\rm{[Hz]}$}
& \multicolumn{1}{c}{\bf Data} \\
&\\
\hline\\
                 & \multicolumn{1}{l}{\bf ${\bf\tau}$+{$\bf e^+e^-$}}
& \multicolumn{1}{l}{$\bf e^+e^-$} 
                 & \multicolumn{1}{l}{\bf ${\bf\tau}$+$\bf e^+e^-$}
& \multicolumn{1}{l}{$\bf e^+e^-$}\\ 
{\bf Light Isovector}&&&\\
$4m_\pi^2\rar 0.8$&$314.5\pm 2.3$&$302.7\pm 7.1$&$152.9\pm 1.8$&$148.4\pm
3.1$&\cite{DAVIE,ALEPH,OPAL}\\
$0.8\rar 2.1$&$77.2\pm 3.4$&$82.0\pm 5.4$&$12.1\pm 0.5$&$16.9\pm 1.9$&\cite{ALEPH,OPAL}\\
$2.1\rar 3.$&$62.3\pm 9.2$&$53.6\pm 4.9$&$7.8\pm 1.2$&$6.7\pm 0.6$&\cite{ALEPH,OPAL}\\
\it Total Light I=1&$\it 454.0\pm 10.6$&$\it 438.2\pm 10.2$&$\it 172.8\pm 2.2$&$\it 172.1\pm 3.7$\\
\multicolumn{1}{l}{\it Average }
& \multicolumn{2}{c}{$\it 446.1\pm 10.4\pm 7.9$}
& \multicolumn{2}{c}{$\it 172.5\pm 3.0\pm 0.3$}
& \multicolumn{1}{c}{} \\
\bf Light Isoscalar&&\\
{\it Below 1.93 }&\\
\multicolumn{1}{l}{$\omega$}
& \multicolumn{2}{c}{$31.5\pm 1.1$}
& \multicolumn{2}{c}{$12.7\pm 0.4$}
& \multicolumn{1}{l}{NWA \cite{PDG}}\\
\multicolumn{1}{l}{$\phi$}
& \multicolumn{2}{c}{$52.3\pm 1.2$}
& \multicolumn{2}{c}{$13.7\pm 0.3$}
& \multicolumn{1}{l}{NWA \cite{PDG}}\\
\multicolumn{1}{l}{$0.66\rar 1.93$}
& \multicolumn{2}{c}{$11.6\pm 3.0$}
& \multicolumn{2}{c}{$2.7\pm 0.7$}
& \multicolumn{1}{l}{$\sum{\rm excl.}$ \cite{DOL}}\\
{\it From 1.93 to 3~} &\\
\multicolumn{1}{l}{$\omega(1.42),~\omega(1.65)$}
& \multicolumn{2}{c}{$9.4\pm 1.4$}
& \multicolumn{2}{c}{$1.2\pm 0.2$}
& \multicolumn{1}{l}{BW \cite{DM2,PDG}}\\
\multicolumn{1}{l}{$\phi(1.68)$}
& \multicolumn{2}{c}{$14.6\pm 4.6$}
& \multicolumn{2}{c}{$1.7\pm .5$}
& \multicolumn{1}{l}{BW \cite{DM2,DM1,PDG}}\\
\multicolumn{1}{l}{\it Total Light I=0}
& \multicolumn{2}{c}{$119.0\pm 5.9$}
& \multicolumn{2}{c}{$32.1\pm 1.0$}
& \multicolumn{1}{c}{}\\
\bf Heavy Isoscalar&&\\
\multicolumn{1}{l}{$J/\psi(1S\rar 4.415)$}
& \multicolumn{2}{c}{$116.3\pm 6.2$}
& \multicolumn{2}{c}{$4.0\pm 0.2$}
& \multicolumn{1}{l}{NWA \cite{PDG}}\\
\multicolumn{1}{l}{$\Upsilon(1S\rar 11.020)$}
& \multicolumn{2}{c}{$12.7\pm 0.5$}
& \multicolumn{2}{c}{$0.1\pm 0.0$}
& \multicolumn{1}{l}{NWA \cite{PDG}}\\
\multicolumn{1}{l}{$T(349)$}
& \multicolumn{2}{c}{$-(0.1\pm 0.0)$}
& \multicolumn{2}{c}{$\approx 0$}
& \multicolumn{1}{l}{NWA \cite{PDG,YND}}\\
\multicolumn{1}{l}{\it Total Heavy I=0}
& \multicolumn{2}{c}{$128.9\pm 6.2$}
& \multicolumn{2}{c}{$4.1\pm .2$}
& \multicolumn{1}{c}{}\\
\bf QCD continuum&&\\
\multicolumn{1}{l}{$3.\rar 4.57^2$}
& \multicolumn{2}{c}{$330.1\pm 1.0$}
& \multicolumn{2}{c}{$17.5\pm.1$}
& \multicolumn{1}{l}{$(u,d,s)$}\\
\multicolumn{1}{l}{$4.57^2\rar 11.27^2$}
& \multicolumn{2}{c}{$503.0\pm 1.0$}
& \multicolumn{2}{c}{$5.0\pm .1$}
& \multicolumn{1}{l}{$(u,d,s,c)$}\\
\multicolumn{1}{l}{$11.27^2\rar (M_Z-3~{\rm GeV})^2$}
& \multicolumn{2}{c}{$2025.7\pm 2.0$}
& \multicolumn{2}{c}{$1.3\pm 0.0$}
& \multicolumn{1}{l}{$(u,d,s,c,b)$}\\
\multicolumn{1}{l}{$(M_Z+3~{\rm GeV})^2\rar 4M_t^2$}
& \multicolumn{2}{c}{$-(794.6\pm 0.6)$}
& \multicolumn{2}{c}{$\approx 0$}
& \multicolumn{1}{c}{$-$}\\
\multicolumn{1}{l}{$Z$-pole}
& \multicolumn{2}{c}{$29.2\pm .5$}
& \multicolumn{2}{c}{$\approx 0$}
& \multicolumn{1}{l}{ppal value \cite{YND}}\\
\multicolumn{1}{l}{$4M_t^2\rar\infty$}
& \multicolumn{2}{c}{$-(24.0\pm 0.1)$}
& \multicolumn{2}{c}{$\approx 0$}
& \multicolumn{1}{l}{$(u,d,s,c,b,t)$}\\
\multicolumn{1}{l}{\it Total QCD Cont.}
& \multicolumn{2}{c}{$2069.4\pm 5.2$}
& \multicolumn{2}{c}{$23.8\pm 1.4$}
& \multicolumn{1}{l}{}\\
\hline
&\\
\multicolumn{1}{l}{\bf Final value}
& \multicolumn{2}{c}{$\bf 2763.4\pm 16.5$}
& \multicolumn{2}{c}{$\bf 232.5\pm 3.2$}
& \multicolumn{1}{l}{}\\
&\\
\hline\\
\end{tabular*}
\end{table*}}
\nin

\begin{figure}[hbt]
\begin{center}
\includegraphics[width=9cm]{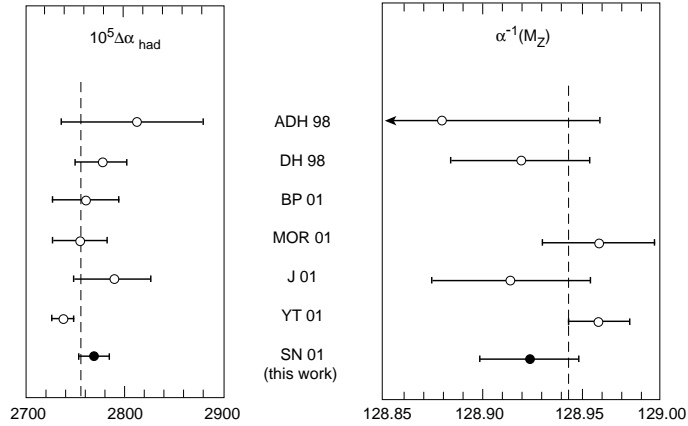}
\caption{Recent determinations of $\Delta\alpha_{\rm had}$ and $\alpha^{-1}(M_Z)$. The dashed vertical
line is the mean central value.}
\end{center}
\end{figure}
\nin 
\subsection{The muonium hyperfine splitting}
Our final result for $\Delta\nu_{\rm had}$ comes from Table 2 of SN2 \cite{SN2}:
\beq\label{nu}
\Delta\nu_{\rm had}=(232.5\pm 3.2)~{\rm Hz}
\eeq
and is shown in Table 3 in comparison with other determinations, where there is an excellent
agreement with the most recent determination \cite{HFS}. Here, due to the $(\ln t)/t$ behaviour of
the kernel function, the contribution of the low-energy region is dominant. However, the $\rho$-meson
region contribution below 1 GeV is 47\% compared with 68\% in the case of $a_\mu$, while the QCD
continuum is about 10\% compared to 7.4\% for $a_\mu$. The accuracy of our result is mainly due to the
use of the
$\tau$-decay data, explaining the similar accuracy of our final result with the one in \cite{HFS} using
new Novosibirsk data. The agreement with \cite{HFS} can be understood from the agreement of
the averaged correlated $e^+e^-$ and $\tau$-decay data compiled in \cite{DAVIE} with the new Novosibirsk
data used in
\cite{HFS}. However, we differ with DH98 \cite{ALFA} in the treatment of the QCD contribution
\footnote{For more details, see \cite{SN}.}. For light quarks,
QCD is applied in the region where non-perturbative contributions are inessential. For heavy quarks, 
perturbative QCD is applied far from heavy quark thresholds, where it can be unambiguously
used.
\begin{table}[hbt]
\begin{center}
\setlength{\tabcolsep}{1.5pc}
\caption{Recent determinations of $\Delta\nu_{\rm had}$}
\label{tab:effluents}
\begin{tabular*}{\textwidth}{@{}l@{\extracolsep{\fill}}ll}
\hline
&\\
Authors&$\Delta\nu_{\rm had}$ [Hz]\\
&\\
\hline
&\\
FKM 99 \cite{FAUST}&$240\pm 7$\\
CEK 01 \cite{HFS}&$233\pm 3$\\
SN 01 \cite{SN2}&$232.5\pm 3.2$\\
&\\
\hline
\end{tabular*}
\end{center}
\end{table}
\nin
\section{Theory of the muon and tau anomalies and new physics}
\subsection{QED and EW contributions to $a_\mu$}
The QED up to 8th order and EW including two-loop
corrections \footnote{References to original works can be found in \cite{SN,JEGER,GM2}.} are:
\bea
a_\mu^{QED}&=&116~ 584~ 705.7(2.9)\times 10^{-11}~,~~~~~~
a_\mu^{EW}=151(4)\times 10^{-11}~.
\eea
\subsection{Higher order hadronic contributions to $a_\mu$}

Higher order hadronic contributions have been discussed first in \cite{CALMET}. They can
be divided into two classes \footnote{For more details of the following discussions, see \cite{SNB}.}. The
first one involves the vacuum polarization and can be related to the measured $e^+e^-\rar$ hadrons total
cross-section, similar to the lowest order contribution.
After rescaling the result in \cite{CALMET,KRAUSE}, one obtains \cite{SN}:
\beq
a_\mu^{had}(h.o.)_{V.P}=-101.2(6.1)\times 10^{-11}~,
\eeq
The second class is the light-by-light scattering diagram.
Contrary to the case of vacuum polarizations, this contribution is not yet fully
related to a direct measurable quantity. In order to estimate this contribution,
one has to introduce some theoretical models. The ones used at present are based on
chiral perturbation \cite{KINO2} and ENJL model \cite{BIJ}. To both are added
vector meson dominance and phenomenological parametrization
of the pion form factors.Ê
The different contributions are summarized in Table \ref{tab: lbyl},
where the first two come from the boson and quark (constituent) loops, while the last
one is due to meson pole exchanges. The first two contributions are quite sensitive to the
effects of rho-meson attached at the three off-shell photon legs which reduce the contributions
by about one order of magnitude (!). The third one with pseudoscalar meson exhanges (anomaly)
gives so far the most important contribution. There is a complete agreement between the two
model estimates (after correcting the sign of the pseudoscalar and axial-vector contributions
\cite{KNECHT}), which may indirectly indicate that the results obtained are model-independent\footnote{See
however
\cite{WISE}.}. Neverthless, there are still some reamining subtle issues to be understood (is the
inclusion of a quark loop a double counting ?, why the inclusion of the rho-meson decreases drastically
the quark and pion loop contributions ? is a single meson dominance justified?..).
The results in Refs. \cite{KINO2} and \cite{BIJ}, after correcting the sign of the pseudoscalar and axial-vector
contributions as suggested in \cite{KNECHT},
are given in the table:
\begin{table*}[H]
\setlength{\tabcolsep}{1.5pc}
\catcode`?=\active \def?{\kern\digitwidth}
\begin{center}
\caption{{$\bf a_\mu^{had}(h.o)_{LL}\times 10^{11}$}}
\label{tab: lbyl}
\begin{tabular*}{\textwidth}{@{}l@{\extracolsep{\fill}}lll}

\hline
&\\
Type of diagrams&{\bf
Ref. \cite{KINO2}}&{\bf Ref. \cite{BIJ} }\\
&&\\
\hline
&&\\
$\pi^-$ loop&$-4.5(8.1)$&$-19(13)$\\
quark loop&9.7(11)&21(3)\\
$\pi^0,\eta,\eta'$ poles&82.7(6.4)&85(13)\\
axial-vector pole&1.74&2.5(1.0)\\
scalar pole&$^{*)}$&$-6.8(2.0)$\\
&\\
\hline
&\\
Total&82.8(15.2)&82.7(18.8)\\
&\\
\hline
\end{tabular*}
\end{center}
{\scriptsize$^{*)}$ We have added here the result from \cite{BIJ}.}

\end{table*}
\nin
while a naive constituent quark model gives \cite{YND} using the result of \cite{LAPORTA}:
\beq
a_\mu^{had}(h.o)_{LL}=92(20)\times 10^{-11}~.
\eeq
Due to the unknown real value of the virtual photon momenta entering into the calculation, it can happen that
none of the previous approaches describes accurately the LL effect \footnote{I thank Eduardo de Rafael and
Paco Yndurain for some clarifying communications on this point.}.  Therefore, for a conservative estimate,
we take an arithmetical average of the three determinations:
\beq
a_\mu^{had}(h.o)_{LL}=84.7(18.0)\times 10^{-11}~.
\eeq
One can notice, for the muon, a strong cancellation between the higher order
vacuum polarisation and the light by light scattering contributions.
\subsection{The total theoretical contributions}
Summing up different contributions, the present theoretical status in the standard
model is \cite{SN}:
\bea
a_\mu^{SM}&=&116~ 584~ 840.2(19.6)\times 10^{-11}+a_\mu^{had}(l.o)\nnb\\
&=&116~591~860.8(78.1)\times
10^{-11}~,
\eea
where $a_\mu^{had}(l.o)$ is the lowest order hadronic contributions evaluated in SN1 \cite{SN} (see Eq.
(\ref{final})). Comparing this SM prediction with the measured value \cite{GM2}:
\beq\label{exp}
a_\mu^{exp}=116~ 592~ 023 (151)\times 10^{-11}~,
\eeq
we deduce:
\beq\label{range}
a_\mu^{new}\equiv a^{exp}_\mu-a^{SM}_\mu=162(170)\times 10^{-11}~.
\eeq
If the future data will be accurate by $\pm 40\times 10^{-11}$, while the theoretical errors are almost
unchanged due to different limitations discussed previously, then the error in $ a_\mu^{new}$ will be reduced
by a factor 2, giving a chance to detect a 2$\sigma$ deviation from SM if any. Combined with the mean value of
existing determinations of
$a_\mu(l.o)^{had}$ given in SN1 \cite{SN}, which gives: $a_\mu^{new}=175(170)\times
10^{-11}$, one can deduce the conservative range: 
\beq
-56\leq a_\mu^{new}\times 10^{11}\leq 393~~(90\%~{\rm CL})~.
\eeq
\subsection{Bounds on some new physics from $a_\mu$}
This result gives, for a supersymmetric model with degenerate sparticle mass, a lower bound of about 113 GeV
\footnote{For more recent detailed discussions, see e.g. \cite{ELLIS,NATH} and references quoted
there.}, while the compositeness and the leptoquark scales are constrained to be above 1 TeV. Bound on
the sparticle mass is comparable with present experimental bound from LEP data. The one of the
leptoquarks is much larger than the present lower bounds of about
$(200\sim 300)$~GeV from direct search experiments at HERA and Tevatron. We expect that these different bounds
will be improved in the near future both from accurate measurements of $a_\mu$ and of $e^+e^-$ data necessary
for reducing the theoretical errors in the determinations of the hadronic contributions, being the major
source of the theoretical uncertainties.
\subsection{Tau anomaly}
In the same way, the higher order hadronic contributions read \cite{SN,SNGM2}
\beq
a_\mu^{had}(h.o)_{VP}=7.6(0.2)\times 10^{-8}~,~~~
a_\mu^{had}(h.o)_{LL}=23.9(5.1)\times 10^{-8}~,
\eeq
which, in the $\tau$ case, the two effect add each others. Adding the other contributions, one obtains \cite{SN}:
\bea\label{atau}
a_\tau^{SM}&=117~759.1(6.7)\times 10^{-8}~.
\eea
This value can be compared with the present (inaccurate) experimental one
\cite{TAYLOR}:
\beq
a_\tau^{exp}=0.004\pm 0.027\pm 0.023~,
\eeq
which, we wish, will be improved in the near future.
\section{The QED running coupling $\alpha(M_Z)$}
To the lowest order hadronic contribution in Eq. (\ref{runalfa}), we add the radiative corrections
 taken by adding the effects of the radiative modes
$\pi^0\gamma,~\eta\gamma,\pi^+\pi^-\gamma,...$. We estimate such effects to be:
\beq
\Delta\alpha_{\rm had}=(6.4\pm 2.7)\times 10^{-5}
\eeq
by taking the largest range spanned by the two estimates in \cite{YND} and \cite{DAVIE}. Using the QED contribution to three-loops \cite{JEGER}:
\beq
\Delta\alpha_{\rm{QED}}=3149.7687\times 10^{-5}~,
\eeq
and the Renormalization Group Evolution of the QED coupling:
\beq
\alpha^{-1}(M_Z)=\alpha^{-1}(0)\Big{[}
1-\Delta\alpha_{\rm{QED}}-\Delta\alpha_{\rm had}\Big{]}~,
\eeq
one obtains the final estimate:
\beq\label{alfa2}
\alpha^{-1}(M_Z)=128.926(25)~,
\eeq 
which we show in Fig 1 for a comparison with recent existing determinations. One can notice
an improved accuracy of the different recent determinations \cite{JEGER,DAVIE,YND,ALFA}. We expect 
that with this new improved estimate of
$\alpha(M_Z)$, present lower bound of 114 GeV from LEP data on the Higgs mass can be improved.
\section{Muonium hyperfine splitting}
Adding to this result in Eq. (\ref{nu}) from SN2 \cite{SN2} , the QED contribution up to fourth order, the
lowest order electroweak contribution
\cite{HFS}, and an estimate of the higher order
weak and hadronic contributions
\cite{POPOV}:
\bea
&&\Delta\nu_{\rm QED}=4~270~819(220)~{\rm Hz}~,
\Delta\nu_{\rm weak}(l.o)=-\frac{G_F}{\sqrt{2}}m_em_\mu\ga\frac{3}{4\pi\alpha}\dr\nu_F
\simeq
-65~{\rm Hz}~,\nnb\\
&&|\Delta\nu_{\rm weak}(h.o)|\approx 0.7~{\rm Hz}~,~~~~~~~~
\Delta\nu_{\rm had}(h.o)\simeq 7(2)~{\rm Hz}~,
\eea
one obtains the Standard Model (SM)  prediction:
\bea
\nu_{\rm SM}\equiv \nu_F+\Delta\nu_{\rm QED}+\Delta\nu_{\rm weak}+\Delta\nu_{\rm had}+
\Delta\nu_{\rm had}(h.o)~.
\eea
If one uses the relation:
\beq\label{nuf}
\nu_F=\rho_F\ga\frac{\mu_\mu}{\mu^e_{B}}\dr \frac{1}{(1+m_e/m_\mu)^3}~:~~~~~~~~~
\rho_F=\frac{16}{3}(Z\alpha)^2 Z^2cR_\infty~,
\eeq
and
$Z=1$ for muonium, $\alpha^{-1}(0)$=137.035 999 58(52) \cite{GM2}, $cR_\infty$ =3 289 841 960 368(25)
kHz
\cite{MOHR}, one would obtain:
\beq
\nu_{\rm SM}=4~463~302~913(511)(34)220)~{\rm Hz}~,
\eeq
 where the two first errors are due to the one of the Fermi splitting energy. The first largest one
being induced by the one of the ratio of the magnetic moments. The third error is due to the 4th order QED
contribution where, one should notice that, unlike the case of $a_\mu$, the dominant errors come from the
QED calculation which should then be improved. This prediction can be compared with the precise data~\cite{EXP}:
\beq
\nu_{\rm exp}=4~463~302~776(51)~{\rm Hz}~.
\eeq
Therefore, at present, we find, it is more informative to extract the Fermi splitting energy $\nu_F$
from a comparison of the Standard Model (SM) prediction with the experimental value of $\nu$. Noting
that
$\nu_F$ enters as an overall factor in the theoretical contributions, one can rescale the previous
values and predict the ratio:
\bea
{\nu_{\rm SM}\over\nu_F}=1.000~957~83(5)~.
\eea
Combining this result with the previous experimental value of $\nu$, one can deduce the
SM prediction:
\beq\label{nf}                   
\nu_F^{SM}=4~459~031~783(226)~{\rm Hz}~,
\eeq
where the error is dominated here by the QED contribution at fourth order.
However, this result is a
factor two more precise than the determination in \cite{HFS} given in Eq. (\ref{eqhfs}), where the
main error  in Eq. (\ref{eqhfs}) comes from the input values of
the magnetic moment ratios. Using this result in Eq. (\ref{nf}) into the expression:
\beq
\nu_F=\rho_F\ga\frac{m_e}{m_\mu}\dr\frac{1}{(1+m_e/m_\mu)^3}\ga
1+a_\mu\dr~,
\eeq
where $\rho_F$ is defined in Eq. (\ref{nuf}),
and $a_\mu=1.165~920~3(15)\times 10^{-3}$ \cite{GM2},  
one can extract a value of the ratio of the muon over the electron mass:
\beq\label{emuon}
\frac{m_\mu}{m_e}=206.768~276(11)~,
\eeq
to be compared with the PDG value $206.768~266(13)$ using the masses in MeV units, and with the one
from
\cite{HFS}:
$206.768~276(24)$. In Ref. \cite{EXP}, an accuracy two times better than the present result has
been also obtained. However, in that case, the errors in the QED contributions may have been
underestimated. After inserting the previous value of
$m_e/m_\mu$ into the alternative (equivalent) relation given in Eq. (\ref{nuf}),
one can deduce the ratio of magnetic moments:
\beq\label{muratio}
\frac{\mu_\mu}{\mu^e_{B}}=4.841~970~47(25)\times 10^{-3}~,
\eeq
compared to the one obtained from the PDG values of $\mu_\mu/\mu_p$ and $\mu_p/\mu^e_B$ \cite{PDG}:
$
{\mu_\mu}/{\mu^e_{B}}=4.841~970~87(14)\times 10^{-3}~.
$
In both applications, the results in Eqs. (\ref{emuon}) and (\ref{muratio}) are in excellent agreement with
the PDG values. These remarkable agreements can give strong constraints to some
contributions beyond the Standard Model and are interesting to be explored. 
\section{Conclusions}
We have evaluated the lowest order hadronic and QCD contributions $a_l^{had}({l.o})$, $\Delta\alpha_{\rm
had}$ and
$\Delta\nu_{\rm had}$ respectively to the anomalous magnetic moment, QED
running coupling and to the muonium hyperfine splitting. Our self-contained results given in Eqs. (\ref{final}),
(\ref{runalfa}) and (\ref{nu}), obtained  within the same strategy and data input, are in excellent agreement with
existing determinations and are quite accurate. We have also revised the estimate
of the light by light scattering contributions to $a_{\mu,\tau}$, and
have explored some phenomenological consequences of these results. 

\end{document}